\newcommand\T{\rule{0pt}{2.6ex}}       
\newcommand\B{\rule[-1.2ex]{0pt}{0pt}} 
\documentclass{aa}  
\usepackage{graphicx}
\usepackage{natbib}
\usepackage{lscape}
\usepackage{subcaption}
\usepackage{multirow}
\usepackage{amssymb}
\bibpunct{(}{)}{;}{a}{}{,} 
\usepackage{txfonts}
\usepackage[colorlinks=true, citecolor=blue,linkcolor=red]{hyperref}
\begin{document}

   \title{NGC\,1068: No change in the mid-IR torus structure despite X-ray variability
          \thanks{Based on observations made with ESO Telescopes at the La Silla Paranal Observatory under programme ID 294.B-5017 and 093.B-0177 and on data obtained from the ESO Science Archive Facility.}
 }

   \author{N. L\'opez-Gonzaga\inst{1} \and D. Asmus \inst{2} \and F. E. Bauer\inst{3,4,5}
   \and K. R. W. Tristram\inst{2} \and L. Burtscher\inst{6} \and A. Marinucci\inst{7} \and \\ G. Matt\inst{7} \and F. A. Harrison \inst{8}}

   \institute{Leiden Observatory, Leiden University, P.O. Box 9513, 2300 RA Leiden,
The Netherlands  \\ \email{nlopez@strw.leidenuniv.nl}
         \and
         European Southern Observatory, Alonso de Córdova 3107, Vitacura, Santiago, Chile
         \and Instituto de Astrof\'isica, Facultad de F\'isica, Pontificia Universidad Cat\'olica de Chile, 306, Santiago, Chile
         \and Millennium Institute of Astrophysics, MAS, Nuncio Monse\~nor S\'otero Sanz 100, Providencia, Santiago de Chile
         \and
         Space Science Institute, 4750 Walnut Street, Suite 205, Boulder, Colorado 80301
         \and 
         Max-Planck-Institut f\"{u}r extraterrestrische Physik, Postfach 1312, Gie\ss enbachstr., 85741 Garching, Germany
         \and
         Dipartimento di Matematica e Fisica, Universit\`{a} degli Studi Roma Tre, via della Vasca Navale 84, 00146 Roma, Italy
         \and
         Cahill Center for Astronomy and Astrophysics, California Institute of Technology, Pasadena, CA, 91125 USA
             }

   \date{Received ;}

  
   \abstract
   {Recent {\it NuSTAR} observations revealed a somewhat unexpected increase in the X-ray flux of the nucleus of NGC\,1068. 
   We expect the infrared emission of the dusty torus to react on the intrinsic changes of the accretion disk.}
   {We aim to investigate the origin of the X-ray variation by investigating the response of the mid-infrared environment.}
   {We obtained single-aperture and interferometric mid-infrared measurements and directly compared the measurements observed before and immediately after the X-ray variations.
   The average correlated and single-aperture fluxes as well as the differential phases were directly compared to detect a possible change in the structure of the nuclear emission on scales of $\sim$2\,pc.}
   {The flux densities and differential phases of the observations before and during the X-ray variation show no significant change over a period of ten years.
   Possible minor variations in the infrared emission are $\lesssim$\,8\,\%.}
   {Our results suggest that the mid-infrared environment of NGC\,1068 has remained unchanged for a decade. 
   The recent transient change in the X-rays did not cause a significant variation in the infrared emission.
   This independent study supports previous conclusions that stated that the X-ray variation detected by {\it NuSTAR} observations is due to X-ray emission piercing through a patchy section of the dusty region. 
   }
   \maketitle
%

   \graphicspath{ {/home/nlopez/Documents/Papers/NGC1068_var/Figures/}  }

\section{Introduction}

The galaxy \object{NGC\,1068} \citep[$D_L=14.4$\,Mpc;][]{1988ngc..book.....T} is typically referred to as a prototype Seyfert type~II galaxy.
It has been intensely studied for many years, providing broad support for a unification theory of active galactic nuclei (AGN) \citep{1993ARA&A..31..473A, 1995PASP..107..803U}.
Early optical polarization observations of NGC\,1068 revealed for the first time the broad-line emission in type~II AGNs and provided evidence of a circumnuclear dusty region that is commonly referred to as the 'torus' \citep{1985ApJ...297..621A, 1991ApJ...378...47M}.

The central engine in AGNs produces X-ray, optical, and UV emission that is partially absorbed and re-emitted in the infrared by circumnuclear dust, which causes a pronounced infrared peak in the spectral energy distribution (SED) of many AGNs \citep{1989ApJ...347...29S}. 
In the case of NGC\,1068, adaptive optics studies with high spatial resolution revealed infrared extended emission \citep{1998ApJ...504L...5B, 2000AJ....120.2904B, 2001ApJ...557..637T, 2005MNRAS.363L...1G}, but the most important step toward resolving the parsec-sized circumnuclear dust structure was achieved with mid-infrared ($\lambda=8$\,--\,$13\,\mu$m) interferometric observations acquired by \citet{2004Natur.429...47J} with the MID-Infrared Interferometric Instrument \citep[MIDI,][]{2003Ap&SS.286...73L} at the European Southern Observatory's (ESO) Very Large Telescope Interferometer (VLTI) located on Cerro Paranal in Chile.
Subsequent efforts to improve the $(u,v)$ coverage \citep{2009MNRAS.394.1325R, 2014A&A...565A..71L} provided the best modeled mid-infrared image of NGC\,1068 so far and allowed us to analyze the structure of the warm dust with great detail.
Based on the modeling reported by \citet{2014A&A...565A..71L}, the mid-infrared environment of NGC\,1068 can be sorted into three distinct components:

\begin{enumerate}
 \item A 1.35$\,\times\,$0.45\,pc hot ($\sim$\,800\,K) component colinear with the H$_2$O megamaser disk  \citep{2009MNRAS.394.1325R}.
 \item A $\sim$\,3$\,\times\,$2\,pc warm ($\sim$\,300\,K) nuclear component considered an extension of the  nuclear hot dust.
 \item A $\sim$\,13$\,\times\,$4\,pc warm ($\sim$\,300\,K) extended component, located  north of the hot nuclear disk, contributing with roughly half of the total 12\,$\mu$m nuclear flux.
\end{enumerate}

In the X-ray regime, the emission of the nuclear region of NGC\,1068 also reveals great complexity. 
NGC\,1068 has been extensively studied at X-ray wavelengths over the past two decades \citep[see, e.g.,][]{1999MNRAS.310...10G, 1997A&A...325L..13M, 2012ApJ...756..180W, 2015ApJ...812..116B} and is considered the best case of a heavily Compton-thick AGN \citep[$N_H$ > 10$^{25}$\,cm$^{-2}$, ][]{2000MNRAS.318..173M}.
Using  {\it NuSTAR} \citep{2013ApJ...770..103H} observations, with unprecedented sensitivity above $\sim$8\,keV and a full energy range of 3\,--\,79\,keV, and previous X-ray data, \citet{2015ApJ...812..116B} found that the observed X-ray emission of NGC\,1068 is consistent with being constant over all past observations at <\,10\,keV (with less than $\sim$\,10\,\% variance) and >\,10\,keV (with less than $\sim$\,30\,\% variance).
The best-fit model that explains the combined {\it NuSTAR}, {\it Chandra}, XMM{\it-Newton}, and {\it Swift}-BAT spectra, spanning a decade in time, is a multi-component reflector of

\begin{enumerate}
 \item a $\approx$\,10$^{25}$\,cm$^{-2}$ nuclear (<\,2\,arcsec) cold reflector consistent with torus reflection,
 \item a $\approx$\,10$^{23}$\,cm$^{-2}$ nuclear cold reflector possibly from tenuous material in the vicinity of photoionized clouds, and
 \item a $\approx$\,5\,$\times$\,10$^{24}$\,cm$^{-2}$ host galaxy (>\,2\,arcsec) cold reflector consistent with distant reflection from large-scale clouds.
\end{enumerate}

\begin{figure}
   \centering
   \includegraphics[width=0.8\hsize]{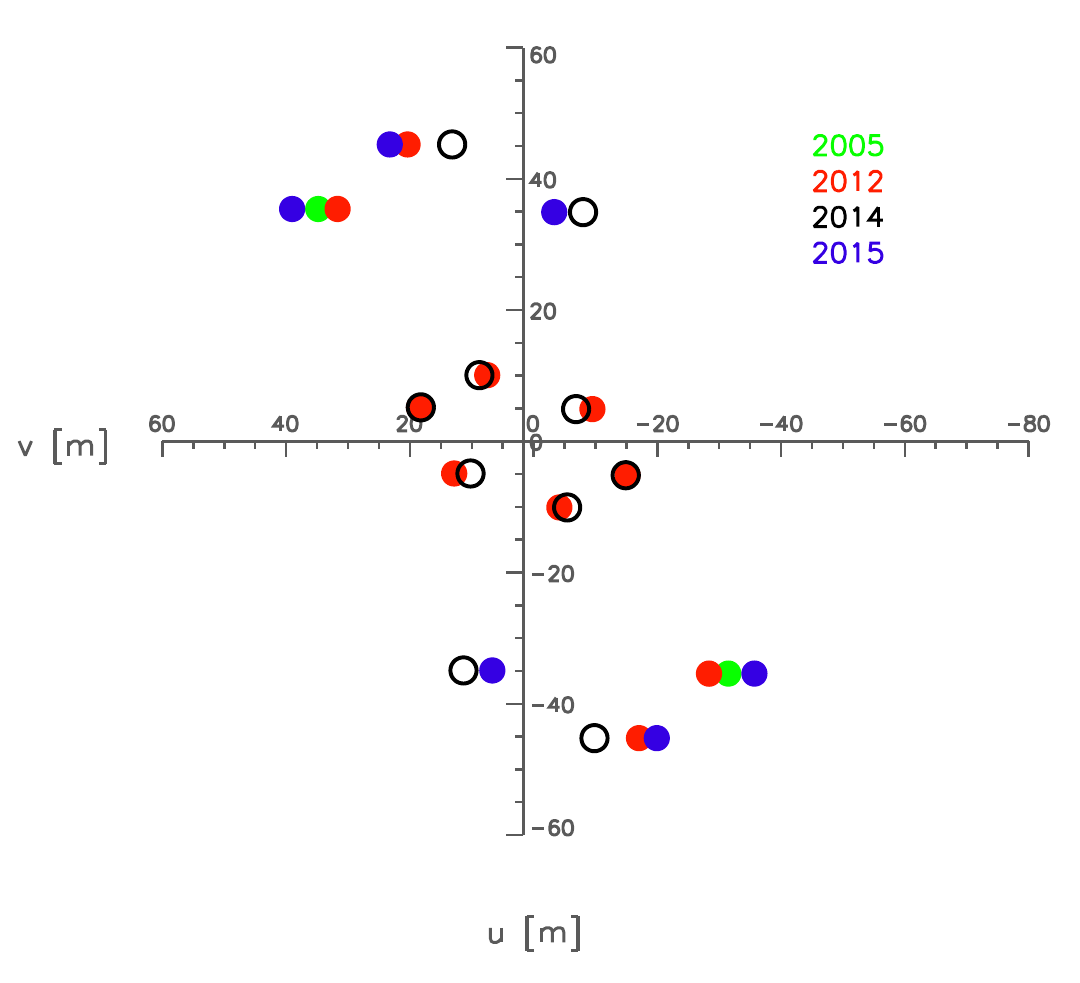}
   \caption{$(u,v)$ coordinates of the data points.
   New $(u,v)$ points acquired during 2014 and 2015 are represented by black open circles and blue filled circles, respectively.
   Previously published interferometric observations obtained in 2005 and 2012 \citep{2009MNRAS.394.1325R, 2014A&A...565A..71L} are shown with green and red circles, respectively. 
   See Sect.~\ref{subsec:int-data} for details on how the data were selected.
   }
   \label{fig:uvplot}
\end{figure}

\subsection{X-ray variability}

Very recently, \citet{2016MNRAS.456L..94M} presented results from a monitoring campaign, between July 2014  and February 2015, using {\it NuSTAR} and XMM{\it-Newton} observations to search for possible variability in the reflection component (Fe K$\alpha$ 6.4\,keV line and $\sim$\,30\,keV Compton hump).
The strength of Fe K$\alpha$ line, measured with the XMM{\it -Newton} data, was found to be constant within 5\,\%. 
However, {\it NuSTAR} observations obtained in August 2014 show a transient excess above 20\,KeV of $32 \pm 6$\,per cent.

The variability was somewhat unexpected, given the model provided by \citet{2015ApJ...812..116B} and previous variability constraints.
According to \citet{2016MNRAS.456L..94M}, the most plausible explanation is a decrease in the total absorbing column of at least $\Delta \mbox{N}_H\simeq2.5$\,$\times$\,10$^{24}\,\mbox{cm}^{-2}$, which permitted the nuclear radiation to pierce the patchy nature of the torus clouds.
Variations lasting for a few tens of days, such as the one observed for NGC\,1068, suggest that the X-ray absorbing clouds detected are most likely dusty \citep{2014MNRAS.439.1403M}. 
This motivated us to search for any infrared response that either confirms the conclusions of \citet{2016MNRAS.456L..94M} or reveals more about this sudden change. 

For an absorption variability effect we do not expect to observe an increase in the mid-infrared flux, as the >\,10\,keV change would be due to shifting clumps of Compton-thick material; in contrast, for a luminosity increase scenario we would expect an increase in the infrared flux or a change in the physical size of the dusty region.
For the latter scenario we should keep in mind that the X-ray burst is not observed directly but rather observed by the increase of the reflected emission from the innermost cold reflecting region. 
Any infrared variation caused by the X-ray burst should therefore be about simultaneous because it is very likely produced in the same region. 
Even for a very short X-ray burst, the reflection emission should live much longer because of the spread in light-crossing times that is due to the extension of the reflection region.
A perfect simultaneity between the X-ray and infrared observations is not necessary.

We expect any variability from the accretion disk to be observed first in the innermost compact component of the dusty environment of NGC\,1068 \citep{2011A&A...534A.121H}. 
Variations in the compact dusty torus might be washed out by the dominating flux of the extended mid-infrared emission of NGC\,1068, and therefore we require high-resolution techniques to isolate the nuclear infrared emission on the torus.
In this paper, we present new mid-infrared single-aperture data and interferometric observations of NGC\,1068, and we investigate possible variations in the mid-infrared emission.
The outline of the paper is as follows. Section~\ref{sec:mid-obs} describes the new observations, data reduction, and calibration. 
The reduced data are analyzed and discussed in Sect.~\ref{sec:results}.
And finally, our conclusions are presented in Sect.~\ref{sec:conclusions}.

\section{Mid-infrared data: Observations, data reduction, and calibration}
\label{sec:mid-obs}

\begin{table}
   \caption{Nuclear single-aperture flux densities ($F_\nu$) for two different epochs.}
   \label{tab:datavisir}
   \centering
   \tiny
   \begin{tabular}{c | c c | c c}
      \hline\hline
      \multirow{1}{*}{NIGHT} & \multicolumn{2}{c |}{FILTERS}   &  \multicolumn{2}{c}{EPOCH RATIOS $(F_{2015}/F_{2005})$} \T\B\\
                 &  ARIII  &  NE\_II &  \multirow{1}{*}{ARIII} & \multirow{1}{*}{NE\_II} \T\\
                 &  (Jy)     &  (Jy)   &    &   \B\\
      
      \hline
      
      2005-11-15 &  $5.73 \pm 1.90$    &  $12.30 \pm 3.22$ &  \multirow{2}{*}{$1.32 \pm 0.70$} & \multirow{2}{*}{$0.97 \pm 0.35$} \T\B\\
      
      2015-01-11 &  $7.57 \pm 3.14$    &  $11.97 \pm 3.02$ & \B\\
      \hline\hline
   \end{tabular}
\end{table}

\subsection{Single-aperture data }

Single-aperture mid-infrared images and spectra were taken with the VLT spectrometer and imager for the mid-infrared \citep[VISIR, ][]{2004Msngr.117...12L}.
VISIR is mounted at the Cassegrain focus of the UT3 telescope of the ESO-VLT on Cerro Paranal, Chile. 
It provides diffraction-limited imaging at high sensitivity in three mid-infrared atmospheric windows (M, N, and Q~band) and features a long-slit spectrometer with a range of spectral resolutions between 150 and 30000.
The data were acquired during a VISIR post-upgrade recommissioning run \citep{2015Msngr.159...15K} for testing purposes on the nights of January 9 and 10, 2015, and were later released to the public.  
Unfortunately, the mid-infrared seeing was not favorable on either night, with a point spread function (PSF) full width at half maximum in the $N$~band of >\,0.4\,arcsec.
On the first night, a low-resolution $N$~band spectrum was recorded with a 0.75\,arcsec slit width, an exposure time of 3\,minutes, and 8\,arcsec chopping along the slit.
It covers the whole spectral region between 8 and 13\,$\mu$m.  
A spectrophotometric standard star from the Cohen catalog \citep{1999AJ....117.1864C} with similar setting was observed immediately after.  
The data were reduced with the ESO pipeline with default parameters, and extraction was performed on the nucleus with a width of 0.44\,arcsec, matched to the seeing. 
The images were taken during the night of January 10, 2015 in the ARIII ($8.99\pm0.14\,\mu$m) and NEII\_2 ($13.04\pm0.22\,\mu$m) filters with an exposure time of 4\,minutes, a chopping throw of 15\,arcsec, and perpendicular nodding.  
A corresponding photometric standard star also from the Cohen catalog was observed immediately before. 
These data were also reduced with the ESO pipeline with default settings, while the photometry of the nuclear emission was performed with the \textsc{mirphot} package from \citet{2014MNRAS.439.1648A}.  

 \begin{table*}
   \caption{Averaged observed interferometric quantities for different baseline configurations. {\it Name:} baseline configuration,  {\it BL:} projected baseline length, {\it P.A.:} position angle, {\it u,v :} ($u,v$) coordinates. 
    Uncertainties delimit the 1$\sigma$ area of the observed measurements. 
  }
   \label{tab:data}
   \centering
   \tiny
   \begin{tabular}{c | c c c c c | c c c c c c | c c }
      \hline\hline
      NIGHT & \multicolumn{5}{c |}{BASELINE INFORMATION} & \multicolumn{6}{c |}{CORRELATED FLUX} & \multicolumn{2}{c}{DIFFERENTIAL PHASE} \T\B\\
         & Name & BL & P.A. & $u$ & $v$ & \multicolumn{3}{c}{Average values [Jy]} &  \multicolumn{3}{c |}{Epoch ratios} & Amplitude & Epoch\\
         &  &  [\,m\,] &  [$\,^\circ\,$] & [\,m\,] &  [\,m\,] & $F_{8.5\,\mu\text{m}}$ &  $F_{10.5\,\mu\text{m}}$ & $F_{12\,\mu\text{m}}$ & ${8.5\,\mu\text{m}}$ &  ${10.5\,\mu\text{m}}$ & ${12\,\mu\text{m}}$  & [$\,^\circ\,$]  & difference [$\,^\circ\,$] \\
      \hline\hline
      2012-09-24 &  B2C1  &  10.5 &   28.3 &  -5.0 &  -9.2 &  4.2 $\pm$ 0.4 &  5.5 $\pm$ 0.3 &  9.8 $\pm$ 0.6 & \multirow{2}{*}{1.0 $\pm$ 0.1} & \multirow{2}{*}{1.0 $\pm$ 0.1} & \multirow{2}{*}{ 1.0 $\pm$ 0.1} &  46.0 $\pm$  9.5 & \multirow{2}{*}{8.0 $\pm$ 15.6} \T\\
      
      {\bf 2014-09-26} &  "     &  11.1 &   33.3 &  -6.1 &  -9.2 &  4.4 $\pm$ 0.2 &  5.6 $\pm$ 0.3 & 10.1 $\pm$ 0.6 & & & &  53.9 $\pm$  6.1 & \B\\
      
      \hline
      2005-11-13 &  U2U3  &  43.3 &   40.9 &  28.4 &  32.5 &  3.6 $\pm$ 0.5 &  1.6 $\pm$ 0.2 &  4.0 $\pm$ 0.3 & \multirow{2}{*}{1.2 $\pm$ 0.1} & \multirow{2}{*}{0.9 $\pm$ 0.1} & \multirow{2}{*}{ 1.0 $\pm$ 0.1} & -42.9 $\pm$  6.8 & \multirow{2}{*}{7.5 $\pm$ 11.7} \T\\

      2012-09-20 &  G1I1  &  41.4 &   38.4 &  25.7 &  32.5 &  3.4 $\pm$ 0.2 &  1.5 $\pm$ 0.1 &  3.9 $\pm$ 0.2  & & & & -59.2 $\pm$ 10.7 & \\

      {\bf 2015-01-10} &   "    &  45.6 &   44.6 &  32.0 &  32.5 &  3.5 $\pm$ 0.2 &  1.4 $\pm$ 0.1 &  4.8 $\pm$ 0.2  & & & & -43.5 $\pm$  5.5 & \B\\

      \hline

      2007-10-07 &  E0G0  &  15.2 &   71.8 &  14.4 &   4.7 &  7.1 $\pm$ 0.9 &  6.5 $\pm$ 0.6 & 11.3 $\pm$ 1.1 & \multirow{2}{*}{0.9 $\pm$ 0.2} & \multirow{2}{*}{0.9 $\pm$ 0.1} & \multirow{2}{*}{ 1.0 $\pm$ 0.2} &  -4.5 $\pm$  7.8 & \multirow{2}{*}{1.7 $\pm$ 13.9} \T\\

      {\bf 2014-09-26} &  A1C1  &  15.0 &   71.5 &  14.2 &   4.7 &  7.1 $\pm$ 0.7 &  6.2 $\pm$ 0.5 & 10.3 $\pm$ 1.0  & & & &  -2.9 $\pm$  6.1 & \B\\

      \hline

      2012-09-19 &  I1K0  &  44.5 &   21.1 & -16.0 & -41.5 &  2.5 $\pm$ 0.3 &  0.9 $\pm$ 0.1 &  3.2 $\pm$ 0.1  & \multirow{2}{*}{1.0 $\pm$ 0.1} & \multirow{2}{*}{1.2 $\pm$ 0.2} & \multirow{2}{*}{ 1.0 $\pm$ 0.2} &  23.1 $\pm$ 19.7 & \multirow{2}{*}{5.7 $\pm$ 24.6} \T\\

      {\bf 2014-11-17} &  "     &  42.6 &   13.4 &  -11.1 & -41.5 &  2.7 $\pm$ 0.5 &  1.1 $\pm$ 0.1 &  3.2 $\pm$ 0.5  & & & &  27.7 $\pm$ 10.7 & \\

      {\bf 2015-01-23} &  "     &  45.4 &   24.0 & -18.5 & -41.5 &  2.9 $\pm$ 0.1 &  1.0 $\pm$ 0.1 &  3.6 $\pm$ 0.1  & & & &   9.9 $\pm$  3.3 & \B\\

      \hline

      2012-09-26 &  A1B2  &  10.0 &  116.9 &  -8.9 &   4.5 &  5.2 $\pm$ 1.1 &  3.1 $\pm$ 0.6 &  7.3 $\pm$ 0.9  & \multirow{2}{*}{1.0 $\pm$ 0.2} & \multirow{2}{*}{1.1 $\pm$ 0.3} & \multirow{2}{*}{ 0.8 $\pm$ 0.3} & -14.8 $\pm$  8.7 & \multirow{2}{*}{5.4 $\pm$ 16.7} \T\\

      {\bf 2014-09-26} &  "     &   8.6 &  121.7 &  -7.3 &   4.5 &  4.4 $\pm$ 0.4 &  3.3 $\pm$ 0.3 &  7.5 $\pm$ 0.4  & & & &  -9.5 $\pm$  8.0 & \B\\

      \hline

      {\bf 2014-09-30} &  H0I1  &  33.1 &  165.5 &   8.3 & -32.0 &  3.4 $\pm$ 0.4 &  1.8 $\pm$ 0.1 &  5.3 $\pm$ 0.3  & \multirow{2}{*}{1.0 $\pm$ 0.1} & \multirow{2}{*}{1.0 $\pm$ 0.1} & \multirow{2}{*}{ 1.0 $\pm$ 0.1} &  30.1 $\pm$  6.6 & \multirow{2}{*}{2.3 $\pm$ 9.3} \T\\

      {\bf 2015-01-20} &  "     &  32.2 &  176.0 &   2.2 & -32.0 &  3.8 $\pm$ 0.4 &  1.7 $\pm$ 0.1 &  5.4 $\pm$ 0.4  & & & &  27.8 $\pm$  2.7 & \B\\

      \hline\hline

   \end{tabular}
\end{table*}

\normalsize

\begin{figure}
   \centering
   \includegraphics[width=\hsize]{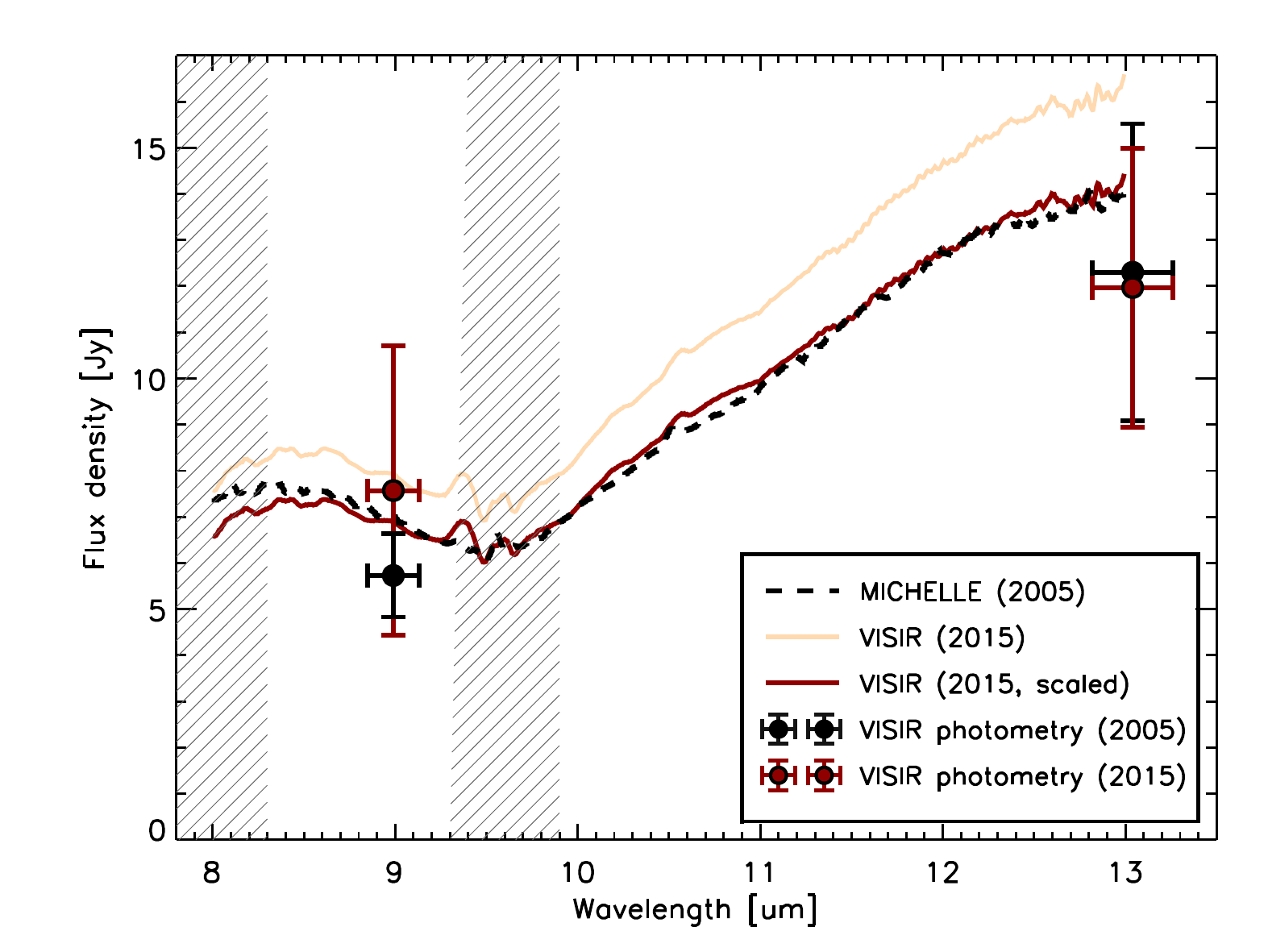}
   \caption{Nuclear mid-infrared spectrophotometry of NGC\,1068 from single-dish observations.
   We include spectra and fluxes from two epochs: before (from 2005) and after (from 2015) the detection of X-ray variability.
   The spectral regions close to 8\,$\mu$m and 9.5\,$\mu$m are affected by strong atmospheric absorption (shaded areas), and these regions should be ignored.
   The spectra have been smoothed to improve readability.
   See Sect~\ref{subsec:visir_obs} for more details.}
   \label{fig:visir}
\end{figure}

\subsection{Interferometric data}
\label{subsec:int-data}

Interferometric measurements were obtained with the instrument MIDI at the ESO's VLTI facility.
The instrument MIDI is a two-beam Michelson interferometer that operates in the $N$~band (8\,--\,13\,$\mu$m) and combines the light from two telescopes; 
a pair of 8.2\,m Unit Telescopes (UTs) or a pair of 1.8\,m Auxiliary Telescopes (ATs). 
The main interferometric observables obtained by MIDI are the correlated flux spectra and the differential phases, which are obtained from the interference pattern generated
by the two beams. 
In the remainder of this paper we use correlated fluxes rather than visibilities. 
The latter are defined as the correlated flux divided by the total or photometric flux.
In the mid-infrared, the difficulties of measuring photometric fluxes against the fluctuations of the bright sky favor the use of correlated fluxes.
The differential phases are identical to the true interferometric phases except that the constant and linear dependencies of phase on wavenumber  $k\equiv 2\pi/\lambda$ have been removed.

Observations with intermediate AT baselines were requested and observed during the nights of January, 10, 20, and 23, 2015 using Director's Discretionary Time (DDT). 
We additionally included published and unpublished interferometric data from our previous campaigns with the requirement that they were observed (nearly) contemporaneously to the period of X-ray variation or observed a few years before. 
These include observations taken on the nights of September, 21, 26, and 30, 2014, and November, 17, 2014, using Guaranteed Time Observations (GTO).
For our observations we used the low-resolution NaCl prism with spectral resolution $R\equiv \lambda/\Delta \lambda \sim 30$ to disperse the light of the beams.
A log of the observations and instrument setup can be found in Appendix~\ref{sec:appendix}.
The published data were taken from \citet{2014A&A...565A..71L}. 
  
We applied the same techniques and method as were used for the MIDI AGN Large Program \citep{2012SPIE.8445E..1GB} to plan our observing strategy and to reduce and analyze the data.

The data were reduced with the interferometric data reduction software {\it MIDI Interactive Analysis and Expert Work Station} \citep[MIA+EWS\footnote{EWS is available for download 
from:\hfil\break http://home.strw.leidenuniv.nl/$\sim$jaffe/ews/index.html.},][]{2004SPIE.5491..715J} , which implements the method of coherent integration for MIDI data.
Calibration of the correlated fluxes was computed by dividing the correlated fluxes of the target by those of the calibrator and multiplying by the known flux of the calibrator. 
For the calibrators HD10380 and HD18322 we used the spectral template of \citet{1999AJ....117.1864C}.
We followed the stacking strategy of \citet{2013A&A...558A.149B}, where fringe tracks on NGC\,1068 are reduced together when they are less than 30\,min apart and are calibrated with the same star.

\begin{figure}
   \centering
   \includegraphics[width=1.1\hsize]{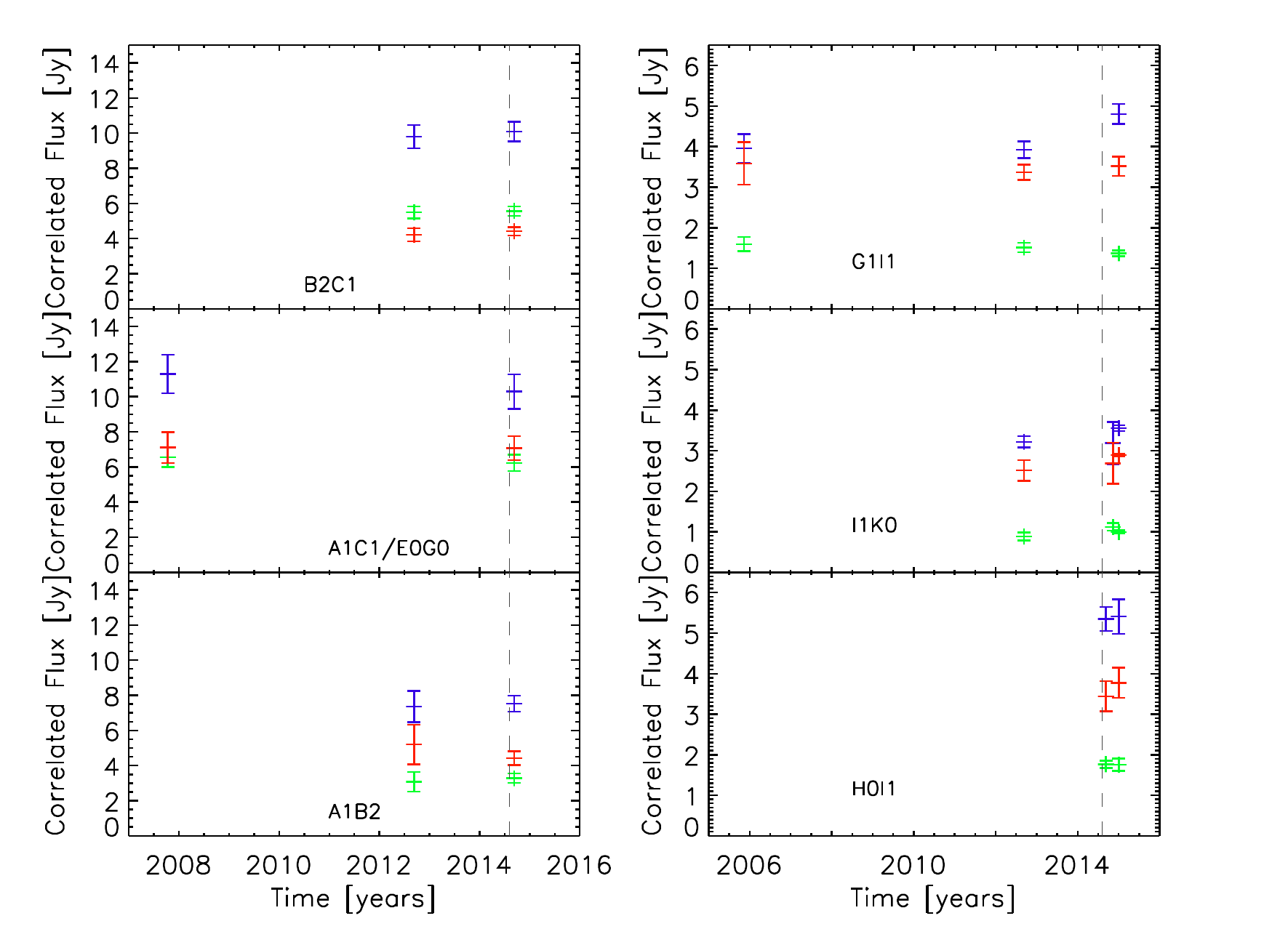}
   \caption{Average correlated fluxes for different epochs grouped by their baseline configuration. The points in each frame have similar ($u,v$) coordinates.
   The color of the symbols indicate the wavelength: Fluxes at 12\,$\mu$m are plotted in blue, those at  10.5\,$\mu$m are shown in green and 8.5\,$\mu$m fluxes in red. The dashed line indicates the time of the reported X-ray variations.}
   \label{fig:var_Figcorr}
\end{figure}

\begin{figure}
   \centering
   \includegraphics[width=\hsize]{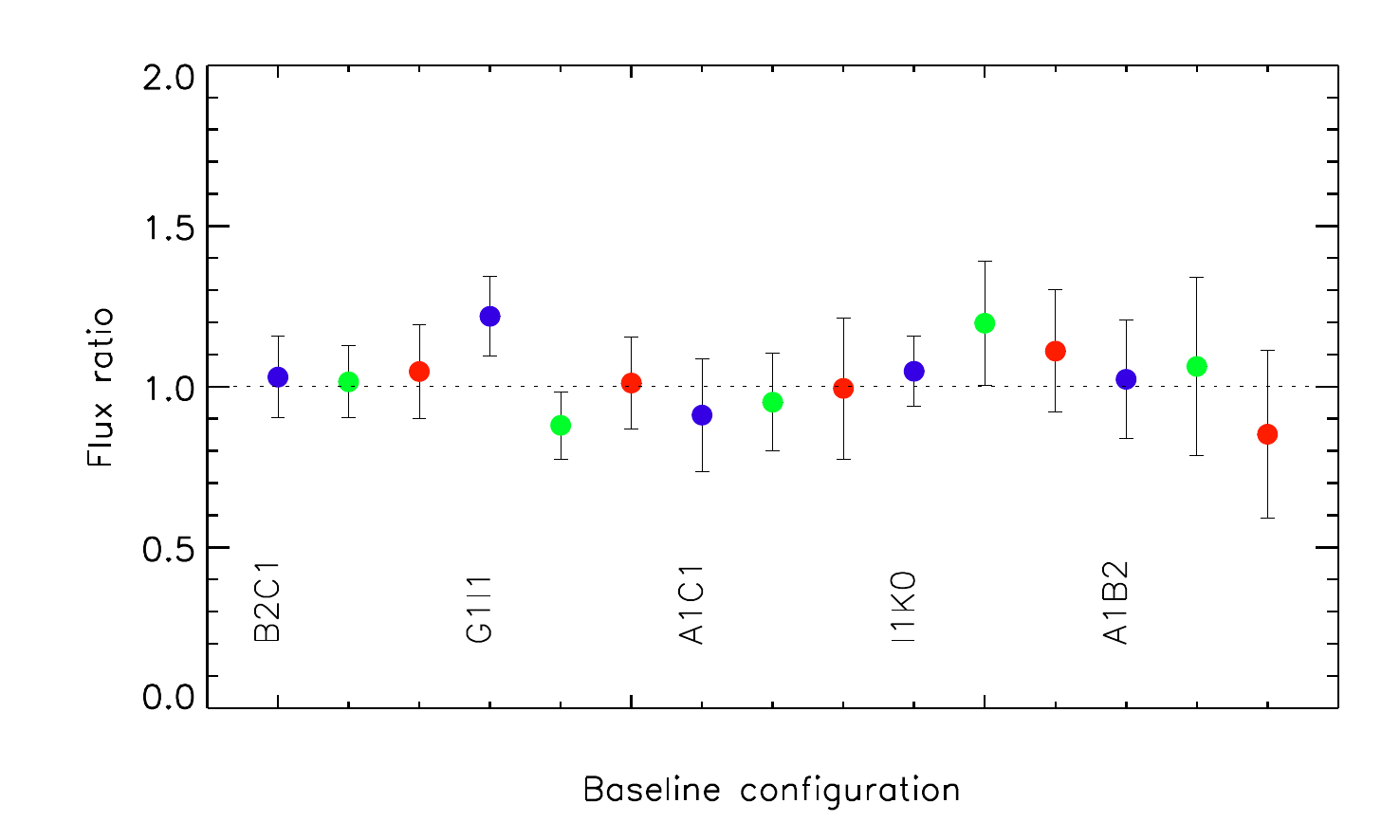}
   \caption{Correlated flux ratios between average measurements from before  and during  the X-ray variations for each baseline and wavelength.
   The color of the symbols is the same as in Fig.~\ref{fig:var_Figcorr}. The dotted line indicates a ratio equal to one, i.e., no variation. }
   \label{fig:flx-ratio}
\end{figure}

\begin{figure}
   \centering
   \includegraphics[width=0.9\hsize]{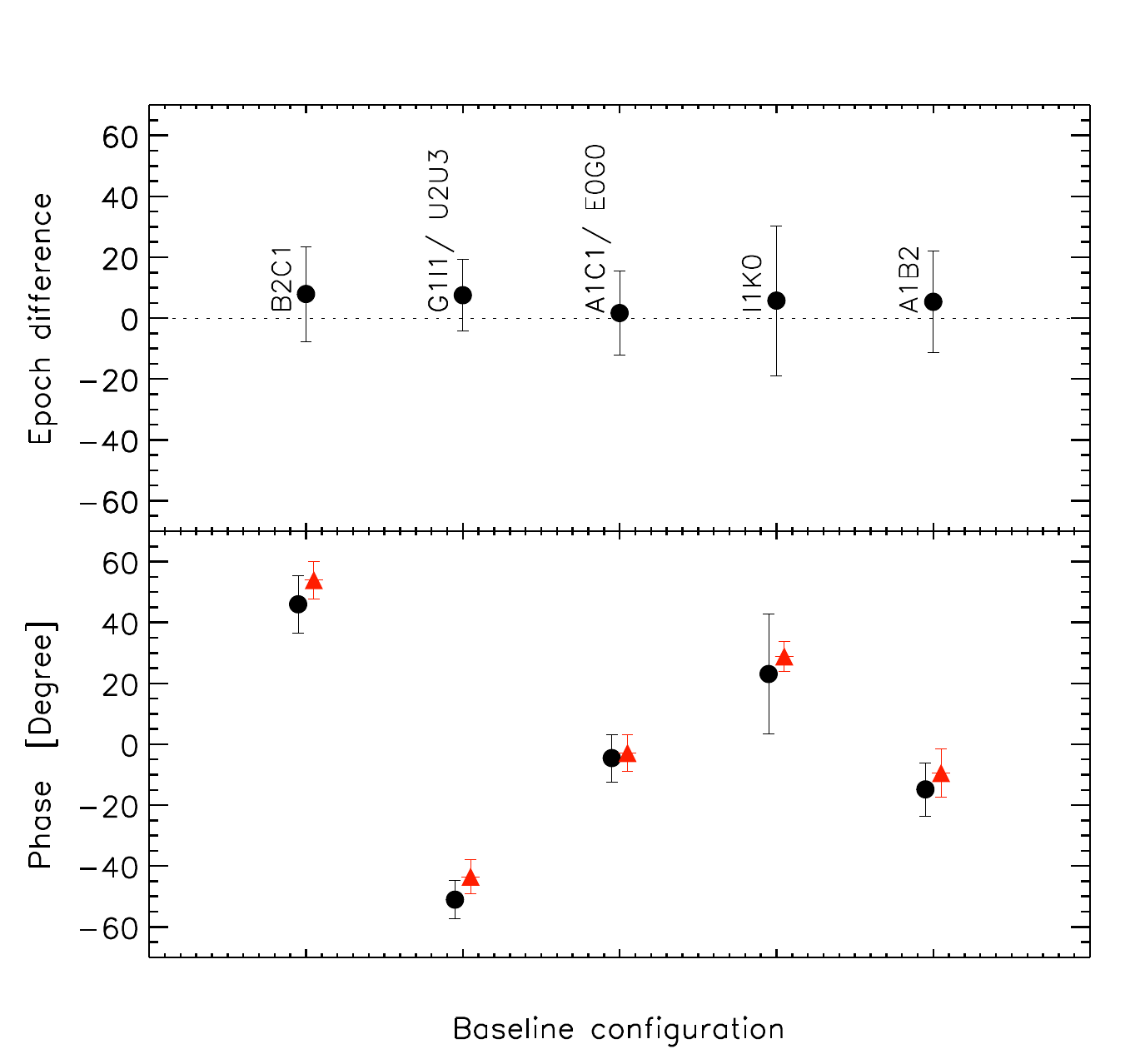}
   \caption{Variations in the amplitude of the differential phase.
   {\it Top)} Differential phase difference between average measurements from before and after the X-ray variation for each baseline configuration.
   The dotted line shows the diference equal to zero. 
   {\it Bottom)} Amplitude of the differential phase before (black circles) and after (red triangles) the X-ray variation for each baseline.}
   \label{fig:phi-ratio}
\end{figure}

\section{Results}
\label{sec:results}

\subsection{VISIR measurements}
\label{subsec:visir_obs}

Extracting the nuclear AGN flux contribution of NGC\,1068 from the single-aperture data is not straight forward because of the bright extended emission. 
The extended dusty emission in NGC\,1068 is heated by the accretion disk, but also by some other heating sources, such as shocks generated from interactions of the jet with the surrounding clouds \citep[see, e.g.,][]{1998ApJ...504L...5B}. 

We estimated the unresolved nuclear flux in the images following \citet{2014MNRAS.439.1648A} with manual PSF scaling, leaving a flat residual for the extended emission.
This allowed us to directly compare the fluxes to historical measurements in the same filters obtained in 2005 that have been analyzed in \citet{2014MNRAS.439.1648A}.  
However, the mid-infrared seeing in the ARIII filter was very unstable and particularly bad during the standard star observation in 2015, which makes the above method infeasible for the measurements in this filter.
To still make a comparison between the historical and new ARIII data possible, we used the theoretical (diffraction-limited) PSF for both epochs, which includes additional sources of uncertainty,
however.
The resulting nuclear flux densities for both filters and both epochs are listed in Table~\ref{tab:datavisir}.  
At $13\,\mu$m, the flux remained constant within the statistical uncertainties between 2005 and 2015.
While at $9\,\mu$m the observed values might seem like a flux increase, the uncertainties on these values are large (up to 40 per cent), and additional systematics arise from using the theoretical PSF for the ARIII. 
The values at $9\,\mu$m are therefore constant within the statistical uncertainties.

The $N$~band spectrum taken on the night before does not show a flux increase at short wavelengths, as shown in Fig.~\ref{fig:visir}.  
For the spectrum, no PSF scaling was performed, leading to the higher flux levels in general compared to the photometry from the images. 
The spectrum is dominated by the well-known silicate absorption feature at $\sim$\,10\,$\mu$m and otherwise is featureless with a red continuum slope.  
The only other $N$~band spectrum of similar spatial resolution that has been published was obtained with Gemini/Michelle in 2005 \citep{2006ApJ...640..612M}.  
It is used for comparison of the VISIR spectrum to an epoch before the X-ray variability event.  
The spectrum was extracted with an aperture of $0.4\times 0.4$\,arcsec also without any PSF-scaling to account for extended emission.  
This explains why it shows slightly higher fluxes than our nuclear photometry from 2005 with VISIR.  
The spectral shapes of the 2005 and 2015 spectra are very similar over the whole range, except for the shortest wavelengths. 
However, atmospheric absorption dominates at these wavelengths, which could explain the difference.  
The general continuum of the 2015 spectrum is higher than the one in 2005, but this is probably caused by the larger extraction area of the 2015 spectrum (0.44\,arcsec).  
If we scale the 2015 spectrum by the ratio of the extraction areas (0.9), then both spectra agree well, as illustrated in Fig.~\ref{fig:visir}. 
This is another indication that the ARIII filter measurement from 2015 is unreliable.
We would also expect the spectrum to have become bluer due to the outburst because the emission from the hotter dust closer to the accretion disk lights up first, while the burst travels outward through the dust distribution.
Therefore, we conclude from the analysis of the single-dish data that no flux change is observed in the mid-infrared on a 20\,parsec scale between 2005 and 2015.

\subsection{MIDI measurements}

In addition to the single-aperture data, we also acquired mid-infrared interferometric measurements.
In total, 19 (stacked) independent ($u,v$) points measured under good weather conditions were reduced and calibrated, plus 17 independent measurements with similar ($u,v$) coordinates to the new data were included from previously published data \citep{2014A&A...565A..71L}. 
 To obtain estimates with a good signal-to-noise ratio of the correlated fluxes and differential phases, we followed the approach of \citet{2014A&A...565A..71L}, where we binned the individual ($u,v$) points when they were within 2\,m in distance for short baseline lengths (5\,--\,15\,m) and within 8\,m\footnote{For intermediate baselines (30\,--\,40\,m) we extended the distance within points up to 8\,m because of the limited amount of observations. This binning is still justified as most of the emission observed at such resolutions comes from the hot compact component (with size 20 $\times$ 6\,mas) that is unresolved or marginally resolved at intermediate baselines. 
\citet{2014A&A...565A..71L} previously showed that the 30\,--\,40\,m baseline ($u,v$) points within 8\,m apart in distance measure always the same spectra within the computed uncertainties. } for intermediate baselines (30\,--\,40\,m). 
We divided all our binned measurements into six groups with similar ($u,v$) coordinates. 
Figure~\ref{fig:uvplot} shows the ($u,v$) coordinates of the grouped data points for different epochs, while the average correlated fluxes as a function of time are shown in Fig.~\ref{fig:var_Figcorr}.
In Table~\ref{tab:data} we include for each group the baseline information, year, average correlated flux at 8.5, 10.5, and 12.0\,$\mu$m, and the mean amplitude of the differential phase. 
The uncertainties reported in this work delimit the 1-$\sigma$ area of the observed measurements.

Because of the limited amount of interferometric data, we cannot model our measurements with complex models or even with Gaussian distributions unless we take many assumptions.
Instead, we performed a direct comparison of the observed quantities on each baseline.
For each baseline and observed quantity ($F_{8\,\mu\text{m}}$, $F_{10.5\,\mu\text{m}}$, $F_{12\,\mu \text{m}}$, and $\Delta \phi$), we computed two distinct average values, one using measurements before the reported increase in the X-rays (before September 2014) and the second one including observations after the increased X-rays (contemporaneous or after September 2014).
We then computed the ratios of the quantities for the two epochs of each baseline with their respective uncertainties using propagation of errors. 
The only ($u,v$) coordinate where we were unable to make a comparison between the two epochs is for the baseline H0--I1, since we only have measurements after the X-ray variation. 
In Fig.~\ref{fig:flx-ratio} we show the ratios for the correlated fluxes for each baseline at different wavelengths. 
Additionally, we show in Fig.~\ref{fig:phi-ratio} the average amplitude of the differential phases, as well as their respective ratios between the two epochs.

As the light takes longer to reach the dust at large scales, we can assume that no changes have occurred so far in the large infrared components of NGC\,1068.
With this assumption, changes in flux should be detected by all the baselines used for this work, as the emission from the hot component is essentially unresolved.
For the same reason, we also do not expect to detect a change in the size of the hot compact component. 
By taking the mean and the standard deviation from all the flux ratios, we obtain an average value for the flux ratio of 1.03 $\pm$ 0.06, 1.02 $\pm$ 0.07, and 0.96 $\pm$ 0.08 for 8.5, 10.5, and 12.0\,$\mu$m, respectively. 
Possible variations in the infrared can be at most 8\% of the total flux, which is a similar percentage as the one measured (10\,\%) by \citet{2014A&A...565A..71L} for the period where NGC\,1068 did not exhibit a change in X-rays.
The differential phases are also consistent with no change over the full measurement period.
The total change in the phases for the two different periods is about $5.6 \pm 7.4\,^\circ$.
The constant differential phases measured during different periods confirm the existence of the phases and also shows that no clear changes have occurred in the past ten years. 
Because of the shutdown of the instrument MIDI, the isolated emission of the dusty torus of NGC\,1068 cannot be further monitored, but any variations arising in the nuclear parsec-scale structure should be detected in the future with the second-generation instrument MATISSE \citep{2008SPIE.7013E..70L}.

\section{Discussion and conclusions}
\label{sec:conclusions}

The constant behavior in the infrared emission of the nuclear region of NGC\,1068 suggests that the observed change in the X-ray regime is probably not due to an intrinsic change in the luminosity of the central accretion disk.
With no significant change in the mid-infrared emission, it is more plausible that the increase in X-ray flux is explained as escaping emission through the patchy nature of the torus clouds or that the nature of the X-ray increase is not the outcome of a thermal variation in the emission of the accretion disk. 
The latter seems less likely to occur because the strength of the X-ray variation in terms of X-ray luminosity seems impossible to explain by any process other than the AGN.

High-resolution mid-infrared observations were obtained and analyzed with the aim of detecting possible variations in the infrared emission of the nuclear dusty region of NGC\,1068. 
Based on the analysis of the single-aperture fluxes, correlated fluxes, and the differential phases, we can conclude that the nuclear mid-infrared environment of NGC\,1068 has remained unchanged for a decade.
Possible variations in the infrared flux are at most about 8\,\% regardless of the activity of the accretion disk (X-ray variability). 
Our results support the idea of \citet{2016MNRAS.456L..94M}, who have stated that it is most likely that the observed flux increase in the X-rays is due to the clumpy nature of the dusty region which the X-ray emission has managed to pierce. 

\begin{acknowledgements}
The authors thank the anonymous referee for the thoughtful and helpful comments.
We also thank the ESO Director's Discretionary Time Committee for accepting the observations of NGC\,1068 few weeks after detecting an increase in the nuclear X-ray flux.
This work was based on observations with ESO telescopes at the La Silla Paranal Observatory under programme ID: 294.B-5017, 093.B-0177, 089.B-0099, 080.B-0928, 076.B-0743.
N. L\'opez-Gonzaga was supported by grant 614.000 from the Nederlandse Organisatie voor Wetenschappelijk Onderzoek and acknowledges support from a CONACyT graduate fellowship.
F. E. Bauer acknowledges support from CONICYT-Chile (Basal-CATA PFB-06/2007, FONDECYT Regular 1141218,  "EMBIGGEN" Anillo ACT1101),
and the Ministry of Economy, Development, and Tourism's Millennium Science Initiative through grant IC120009, awarded to The Millennium Institute of Astrophysics, MAS.
L. Burtscher is supported by a DFG grant within the SPP 1573 ''Physics of the interstellar medium''.
\end{acknowledgements}

\bibliographystyle{aa} 
\bibliography{/home/nlopez/Documents/PhD_thesis/Template_Silv/BackMatter/Bibliography/bib/Bib_silvia}

\begin{appendix}
\section{Log of observations}
\label{sec:appendix}

\begin{table}
\centering
\tiny
\caption{Log of observations: {NGC\,1068}. The columns are the {\it time} of 
fringe track observation; {\it St:} Stacked with the following observation (yes:1, no:0); {\it BL:} projected baseline length; {\it PA:} position 
angle; {\it Air:} airmass of fringe track; {\it OK?:} Goodness of observation (good:1, bad:0); {\it Gframes:} number of good frames; {\it Caltime:} time of the calibrator fringe track observation; $\Delta$\,am: difference in airmass with calibrator.}
\begin{tabular}{c c c c c c c c c}
\hline\hline
Time & St & BL & PA & Air  & OK? & Gframes & Caltime & $\Delta$ am \\
& & [m] & [$^\circ$] &  &  &  &  &  \\

\hline\hline
\multicolumn{9}{l}{2012-09-20: I1K0} \T\\
08:42:47 & 1 &  44 &  20  & 1.2  &  1 &  6402 &   08:50:09 &      0.1\\
09:03:58 & 0 &  44 &  22  & 1.2  &  1 &  5918 &   08:50:09 &      0.1\\
09:07:27 & 1 &  44 &  22  & 1.2  &  1 &  5075 &   09:18:18 &      0.1\\
09:11:03 & 0 &  44 &  22  & 1.2  &  1 &  5690 &   09:18:18 &      0.1\\
\hline
\multicolumn{9}{l}{2014-09-26: B2C1} \T\\
05:11:03 & 0 &   9 &  14  & 1.2 &  1 &  6094 &   05:16:05 &      0.1\\
05:23:14 & 1 &   9 &  16  & 1.2 &  1 & 11442 &   05:29:53 &      0.1\\
05:35:11 & 0 &   9 &  18  & 1.2 &  1 & 11472 &   05:29:53 &      0.0\\
\hline
\multicolumn{9}{l}{2014-09-26: A1C1} \T\\
06:06:26 & 1 &  14 &  71  & 1.1 &  1 & 11459 &   06:13:05 &      0.1\\
06:18:14 & 0 &  15 &  71  & 1.1 &  1 & 11505 &   06:13:05 &      0.0\\
\hline
\multicolumn{9}{l}{2014-09-26: B2C1} \T\\
08:17:43 & 1 &  11 &  33  & 1.2 &  1 &  7438 &   08:24:04 &      0.1\\
08:29:31 & 0 &  11 &  33  & 1.2 &  1 & 11525 &   08:24:04 &      0.1\\
\hline
\multicolumn{9}{l}{2014-09-26: A1B2} \T\\
08:56:40 & 0 &   9 & 119  & 1.3 &  1 & 11385 &   08:51:02 &      0.1\\
09:08:35 & 0 &   8 & 120  & 1.3 &  1 & 11332 &   09:03:05 &      0.1\\
09:37:40 & 0 &   7 & 124  & 1.4 &  1 &  5684 &   09:25:12 &      0.2\\
\hline
\multicolumn{9}{l}{2014-09-30: H0I1} \T\\
09:03:30 & 0 &  33 & 165  & 1.3 &  1 &  7335 &   08:55:52 &      0.2\\
09:07:37 & 1 &  33 & 165  & 1.4 &  1 & 11267 &   09:15:35 &      0.1\\
09:24:03 & 0 &  32 & 168  & 1.4 &  1 & 14392 &   09:15:35 &      0.2\\
09:44:28 & 0 &  32 & 172  & 1.5 &  0 &  1678 &   09:37:27 &      0.2\\
\hline
\multicolumn{9}{l}{2014-11-17: I1K0} \T\\
03:28:04 & 0 &  42 &  12  & 1.1 &  1 &  6623 &   03:19:57 &      0.1\\
03:31:52 & 0 &  42 &  12  & 1.1 &  1 &  6723 &   03:40:02 &      0.1\\
03:54:32 & 1 &  42 &  15  & 1.1 &  1 &  6855 &   04:06:28 &      0.1\\
03:58:21 & 1 &  43 &  15  & 1.1 &  1 &  5963 &   04:06:28 &      0.1\\
04:14:09 & 0 &  43 &  17  & 1.1 &  1 &  3817 &   04:06:28 &      0.1\\
\hline
\multicolumn{9}{l}{2015-01-10: G1I1} \T\\
02:00:53 & 0 &  46 &  45  & 1.3 &  0 &    -1 &   01:45:36 &      0.1\\
02:11:32 & 0 &  46 &  45  & 1.3 &  1 & 11255 &   02:20:01 &      0.1\\
03:22:15 & 1 &  44 &  43  & 1.7 &  1 & 11406 &   03:34:07 &      0.2\\
03:49:18 & 0 &  43 &  41  & 2.0 &  1 &  9336 &   03:34:07 &      0.4\\
\hline
\multicolumn{9}{l}{2015-01-11: G1I1} \T\\
02:22:06 & 0 &  46 &  45  & 1.4 &  0 &  2484 &   02:05:42 &      0.2\\
\hline
\multicolumn{9}{l}{2015-01-20: D0H0} \T\\
00:56:01 & 0 &  62 &  72  & 1.2 &  0 &    -1 &   01:40:14 &      0.0\\
\hline
\multicolumn{9}{l}{2015-01-20: H0I1} \T\\
02:22:37 & 0 &  32 & 172  & 1.5 &  1 &  9407 &   02:05:21 &      0.3\\
02:59:54 & 0 &  32 & 179  & 1.9 &  1 &  9872 &   02:41:20 &      0.4\\
\hline
\multicolumn{9}{l}{2015-01-23: G1I1} \T\\
00:36:39 & 0 &  46 &  45  & 1.2 & 0 &    - &         - &   -\\
00:53:51 & 0 &  46 &  45  & 1.2 & 0 &    - &         - &   -\\
\hline
\multicolumn{9}{l}{2015-01-23: I1K0} \T\\
01:24:31 & 0 &  45 &  24  & 1.3 &  1 & 11348 &   01:15:33 &      0.2\\
\hline
\end{tabular}
\end{table}

\end{appendix}

\end{document}